
\magnification=1200
\hsize 15true cm \hoffset=0.5true cm
\vsize 23true cm
\baselineskip=15pt

\nopagenumbers

\font\grande=cmr10 scaled \magstep4
\font\medio=cmr10 scaled \magstep2
\outer\def\beginsection#1\par{\medbreak\bigskip
      \message{#1}\leftline{\bf#1}\nobreak\medskip\vskip-\parskip
      \noindent}

\def \me {\buildrel <\over \sim}
\def \Me {\buildrel >\over \sim}
\def \pa {\partial}
\def \ra {\rightarrow}
\def \big {\bigtriangledown}
\def \fb {\overline \phi}
\def \rb {\overline \rho}
\def \pb {\overline p}
\def \pr {\prime}
\def \se {\prime \prime}
\def \ti {\tilde}
\def \la {\lambda}
\def \La {\Lambda}
\def \Da {\Delta}
\def \b {\beta}
\def \a {\alpha}
\def \ap {\alpha^{\prime}}

\def \Ga {\Gamma}
\def \ga {\gamma}
\def \sg {\sigma}
\def \da {\delta}
\def \ep {\epsilon}
\def \r {\rho}
\def \om {\omega}
\def \Om {\Omega}
\def \noi {\noindent}

\def\sqr#1#2{{\vcenter{\hrule height.#2pt\hbox{\vrule width.#2pt
height#1pt \kern#1pt\vrule width.#2pt}\hrule height.#2pt}}}

\def\lsim{\mathrel{\rlap{\lower4pt\hbox{\hskip1pt$\sim$}}
    \raise1pt\hbox{$<$}}}         
\def\gsim{\mathrel{\rlap{\lower4pt\hbox{\hskip1pt$\sim$}}
    \raise1pt\hbox{$>$}}}         

\line{\hfil CERN-TH.6572/92}
\line{\hfil DFTT-51/92}
\vskip 2 cm
\centerline {\grande  Pre-Big-Bang}
\vskip 0.5 true cm
\centerline{\grande In String Cosmology}
\vskip 1true cm
\centerline{M.Gasperini}
\centerline{\it Dipartimento di Fisica Teorica dell'Universit\`a,}
\centerline{\it Via P.Giuria 1, 10125 Torino, Italy,}
\centerline{\it Istituto Nazionale di Fisica Nucleare, Sezione di Torino}
\centerline{and}
\centerline{G.Veneziano}
\centerline{\it Theory Division, Cern, Geneva, Switzerland}

\vskip 2 cm
\centerline{\medio Abstract}

\noindent
The  duality-type symmetries of string cosmology naturally lead us
to expect a  pre-big-bang phase of accelerated evolution
as the dual counterpart of
the decelerating expansion era of standard cosmology.
Several properties of this scenario are discussed, including the
possibility that it avoids the initial singularity
and that it provides a large amount of inflation.
 We also discuss how
  possible    tracks of the pre-big-bang era may be looked
for directly in the spectral and "squeezing" properties of
relic gravitons and, indirectly, in   the  distorsion they induce
on the cosmic microwave background.
\noindent
\vskip 1 cm
\noindent
CERN-TH.6572/92

\noindent
July 1992
\vfill\eject

\footline={\hss\rm\folio\hss}
\pageno=1

{\bf 1. Introduction}

The standard cosmological model (SCM) provides
the best account so far for
the observed properties of our present Universe. Nevertheless,
 in spite
of its success, the SCM cannot be extended backward in time  to
arbitrarily large curvatures or temperatures  without running
into   problems.
 We are referring here not only
 to the
well-known phenomenological problems
 (e.g. horizon, flatness, entropy),
 but also
to what is probably the main conceptual difficulty of the SCM,
 the initial singularity.
 The aim of this paper is to present a possible alternative to the
initial singularity, which arises naturally in a string
cosmology context because of the "large scale-small scale" symmetry
(target-space duality)
typical of string theory (see Refs. [1--4] for previous pioneer work along
this line).

What we shall refer to  here   as
 "{\it string cosmology}" is just the string-theory analogue
of the usual Einstein-Friedman (EF) equations of the SCM.
The string-cosmology equations are thus
  obtained by varying the low-energy string theory effective action
-describing the massless background fields-
supplemented with a
phenomenological source term that accounts for possible
 additional bulk string matter at the classical level.
As long as the equation of state of
the classical sources [5] is compatible with the properties
 of string propagation in the massless backgrounds, the resulting
system of coupled equations
enjoys a much larger symmetry than that of standard cosmology [6].

 This definition
of string cosmology is probably insufficient
for a truly "stringy" description of the Universe in the high
 curvature
regime (near the Planck scale), where higher-order
string corrections cannot be neglected, but it is certainly enough to
suggest
the alternative to the singularity of the SCM discussed here.

In the SCM,
curvature, source energy density, and temperature all grow
 as we go backward in time: eventually, they blow up
at the initial
singularity, the so-called big bang.
 A possible way out of this conclusion is to
postulate that the standard picture of  space-time as a smooth
manifold does not survive near and above the Planck scale,
where some
drastic modification of the classic geometry is to be introduced
(see
[7] for recent speculations on this subject). There are also,
however, more conventional alternatives to the standard  singular
scenario. The two simplest
possibilities are illustrated in {\bf Fig. 1}. In
the first case
the curvature stops its growth after reaching a maximal value.
In the second case,
the curvature grows to a maximum value and then starts
decreasing again as we go backward in time.

The first
case corresponds to a de Sitter-like primordial inflationary
phase with unlimited
past extension. It is constrained by phenomenological bounds
(based, as we shall see in Section 4, on the properties of the relic
graviton spectrum), which imply for the final curvature scale a
value at
least four orders of magnitude below the Planck scale. This may
seem
unnatural if one believes that the curvature becomes stable
just because
of quantum gravity effects.
 Moreover, it has been recently argued
[8] that eternal exponential expansion with no beginning is
impossible in the context of the standard inflationary scenario.
If this is so,  the hypothesis of a primordial phase of constant
curvature does not help to avoid the problem of the initial singularity.

The second possibility describes the alternative to the singularity
suggested by the symmetries of string cosmology. In this  case, as shown
in Section 4, the quoted
phenomenological constraints can be evaded  and the curvature
can reach  the Planck scale.
In this context the big-bang does no   longer corresponds
to a singularity, but to an instant of maximal curvature marking the
transition from a  "string-driven" growing-curvature regime (in the
sense explained in Section 2) to the
decelerated evolution of the standard
scenario. We shall call, in general, {\it "pre-big-bang"} this new
phase characterized  by  growing curvature and accelerated evolution
(as one goes forward in time).

It is important to stress that the growth of  curvature,
during such
a phase, may be due not only (and not necessarily) to the
contraction of
some internal space, but also to the superinflationary
expansion [9] of
the three physical spatial dimensions. Thus this scenario is
something
genuinely different from the old oscillating cosmological
model, in
which the present expansion is preceded by a contracting phase.
Indeed,  in this scenario, it is also
possible  for the Universe to be monotonically
expanding (see Section 5). We also stress that this picture is by no means
to be regarded as an
alternative to the standard (even inflationary) cosmology.
It is only a
completion of the standard picture, which cannot be extended
 beyond the
Planck era.

This paper is devoted to the discussion of various aspects of this
possible alternative to the singularity of the SCM, and  is organized as
follows:

 In Section 2 we shall discuss two string-theoretic
motivations
 for the pre-big-bang scenario:   one is based on the
solution of the string equations of
motion in a background with event horizons [5]; the second
rests    on
a property of the low-energy string effective action, called
scale-factor-duality [10--12], a particular case of
a more general
$O(d,d)$-covariance of the string cosmology equations [6,13,14].

In Section 3 a simple example of a string-driven pre-big-bang
 scenario is
used to illustrate the behaviour of the density and of
the temperature
during the phase of growing curvature, and to discuss its possible
relevance as a model of inflation.

If such a scenario is
taken seriously, the important question to ask is whether
observable tracks of the pre-big-bang era may survive and be
available to present observations [cf. the observable
tracks of the big bang, the electromagnetic
cosmic microwave background (CMB)].
 In Section 4 we  show that these tracks may exist, and that they
 are to
be looked for in the relic graviton background, either
directly  or through the distortions they
induce on the CMB, perhaps not an
impossible dream for the not-too-distant future.

In Section 5 we present  solutions to the string cosmology
equations
which interpolate smoothly between the growing-curvature and the
decreasing-curvature phases and in which both the curvature
and the effective
string
coupling (the dilaton) are everywhere bounded. Although these solutions
 do not yet incorporate  a completely self-consistent equation of
state for the string sources, they
represent good candidates for describing the background dynamics
 around the maximum curvature regime.

 Our main conclusions are summarized in Section 6.

\vskip 1.5 cm
{\bf 2. String theory motivations for the pre-big-bang scenario}
\vskip 0.5 cm
\noi
{\bf 2.1 String propagation in a background with shrinking event
horizons}

One argument supporting the pre-big-bang scenario is provided by the
solution to the string equations of motion in a background manifold
with event horizons [5]. Indeed, the curvature growth and the accelerated
evolution, typical of a
pre-big-bang epoch, are
necessarily associated with event horizons whose proper
size decreases in time. This has important
consequences for objects of finite proper size, such as strings.

In order to present this argument, it is convenient to start by recalling
some elementary, kinematical properties of Friedmann-Robertson-Walker
 spatially flat  metrics:
$$
g_{\mu\nu}= diag (1, -a^2(t)\da_{ij}).  \; \eqno(2.1)
$$
$a$ is the scale factor, and $H=\dot a/a$ is the usual Hubble
parameter, where a dot denotes differentiation with respect to the
cosmic time $t$.

Consider the proper distance along a null geodesic
$$
d(t)=a(t)\int_{t_1}^{t_2} dt' a^{-1}(t') . \; \eqno(2.2)
$$
In the limit $t_2 \ra t_{Max}$, where $t_{Max}$ is the maximal future
extension of the cosmic time coordinate on the given
 background manifold,
  $d(t_1)$ defines the proper size of the {\it event horizon} at the
time $t_1$ (i.e. the maximal size of the space-time region within
which a
causal connection can be established). In the limit $t_1 \ra t_{min}$,
where $t_{min}$ is the maximal past extension of the cosmic time
coordinate, $d(t_2)$ defines instead the {\it particle horizon} at the
time $t_2$ (i.e. the maximal portion of space which can be included
inside the past light cone of a given observer).

By means of these definitions, and by using $|H|$ as an indicator of the
curvature scale, it is possible to provide a sort of "causal"
classification of the isotropic and homogeneous backgrounds based on
their asymptotic behaviour, which is reported in {\bf Table I} for the
expanding case, and in {\bf Table II} for the contracting one. Note
that, in both tables, the linear evolution of the horizon is referred to
the cosmic time coordinate; note also the "dual" symmetry which
exchanges particle with event horizon when passing from expanding to
contracting backgrounds.

For the purpose of this paper, three points have to be stressed. The
first point is that we are in presence of event horizons only if
$$
sign \{\dot a\} = sign \{\ddot a\}, \; \eqno(2.3)
$$
that is only in the case of accelerated evolution. Inflation, on the
other hand, is accelerated expansion, and inflation plus the
simultaneous shrinking of some internal dimension may be self-sustained
(in the absence of dominant vacuum contributions) only in the case of
accelerated contraction [15]. Event horizons thus appear naturally
during a phase of dimensional decoupling.

The second point to be stressed is that, for an accelerated evolution,
the proper size of the regions that are initially in causal contact
tends to evolve in time, asymptotically, like the scale factor $a(t)$.
[This can be easily seen from eq. (2.2) by considering a causally
connected region of proper initial size $d(t_1)=t_1$, in a background
which is, for example, in accelerated expansion, i.e.
$a \sim t^\a$, $\a>1$
for $t\ra \infty$. One then finds, for $t_2>>t_1$, that $d(t_2)\simeq
d(t_1)a(t_2)/a(t_1)$]. Comparing this behaviour with the time evolution
of the event horizon, we may thus conclude that the regions which are
initially in causal contact grow faster that the horizon in the case of
accelerated expansion, and contract more
slowly than the horizon in the case of
accelerated contraction. In both cases, they will tend to cross the
horizon. This is not a source of problems for point-like objects, of
course, but extended sources may become, in such a situation, larger
than the causal horizon itself.

The third important point is that, according to Tables I and II, the
curvature is growing only if
$$
sign \{\dot a\} = sign \{\dot H\}. \; \eqno(2.4)
$$
A typical pre-big-bang configuration with accelerated evolution and
growing curvature is thus necessarily associated with the presence of an
event horizon which shrinks linearly with respect to cosmic time:
extended objects, in such backgrounds, are doomed to become causally
disconnected, quite independently of their proper size.

This effect is peculiar to gravity, and can be easily understood in
terms of tidal forces [16]. Consider indeed a free-falling object, of
finite size $\la$, embedded in a cosmological background. Each point of
the extended object will fall, in general, along different geodesics. In
the free-falling frame of one end of the object, the other end, at a
proper distance $|z|=\la$, will have a relative acceleration given by
the equation of geodesic deviation,
$$
|{Du^\mu \over Ds}|=|R^\mu\,_{\nu \a \b } u^\nu u^\a z^\b|=\la |{\ddot
a\over a}|. \; \eqno(2.5)
$$
This acceleration defines a local Rindler horizon at a distance
$$
d=\la ^{-1}|{\ddot a \over a}|^{-1}, \; \eqno(2.6)
$$
which depends only on the background curvature. If the curvature is
growing, this distance becomes smaller and smaller, until the two ends
become causally disconnected.

In the case of strings, the existence of a causally disconnected
asymptotic regime implies that an approximate description of the string
motion around the classical path of a point particle is no longer
allowed [17]. Suppose indeed that the exact solution of the string
equations of motion and constraints  in a cosmological background is
expanded
around a comoving geodesic representing the point-particle motion of the
centre of mass of the string [18]. Such an expansion is valid, provided
[5,17]
$$
|\dot X^i| \sim |{\pa X^i\over \pa \sg}|\eqno(2.7)
$$
($X^i$ is any spatial component of the target space coordinates, and
$\sg$ is the usual world-sheet space variable) and provided the
first-order fluctuations, whose Fourier components satisfy the
linearized equation
$$
\ddot \chi^i_n +({n^2\over \la^2}-{\ddot a\over a})\chi^i_n=0,
 \eqno(2.8)
$$
stay small [$\chi^i=a(t)X^i$, and $\la=m\ap$, where $m$ is the string
mass and $\ap$ the string tension].

If the background curvature is growing, this approximation breaks down
asymptotically. Indeed, in the high curvature limit $|\ddot a/a|>>n^2\la
^2$, the asymptotic solution to the fluctuation equation (2.8) can be
written in general as
$$
X^i=A + B \int {dt\over a^2} \eqno(2.9)
$$
  with $A$ and $B$ independent of time, and one obtains
$$
|\dot X^i|<<|{\pa X^i \over \pa \sg}|\eqno(2.10)
$$
in the case of expanding backgrounds, and
$$
|\dot X^i|>>|{\pa X^i \over \pa \sg}|\eqno(2.11)
$$
in the contracting case.

In both cases the condition (2.7) is not satisfied, and the geodesic
expansion is no longer valid, asymptotically. In both cases,
however, the exact solution can be expanded around a new asymptotic
configuration [5,19,20], which is "unstable", in the sense that
it is non-oscillating with respect to the world-sheet time.

What is important, in our context, is that this kind of solution,
when inserted into
the string stress tensor, leads in general to a self-sustained
scenario. In other words it leads to an effective equation of
state characterizing a source which can drive, by itself, the phase of
accelerated evolution and growing curvature. In the perfect fluid
approximation and, in particular, for a background with $d$ expanding
and $n$ contracting dimensions, such an equation of state takes the form
[5]
$$
\r + dp-nq=0\eqno(2.12)
$$
($\r$ is the energy density, $p$ and $q$ are respectively the pressure
in the expanding space and in the internal shrinking dimensions).

According to this equation of state
it is possible to obtain a "string-driven" pre-big-bang phase not
only during an anisotropic situation of dimensional decoupling [5], but
also in the case of isotropic expansion ($n=0$) or isotropic contraction
($d=0$), provided the correct string gravity equations, including the
dilaton, are used instead of the Einstein equations [6]. This is to be
contrasted with the case of point-like sources, where it is impossible to
achieve, asymptotically, the causally disconnected regime characterized
by non-oscillating configurations, and it is thus impossible to arrange
a self-consistent equation of state able to sustain the pre-big-bang
phase.
\vskip 1 cm
{\bf 2.2 Scale factor duality}

A second, probably stronger, motivation supporting the pre-big-bang
picture follows from a property of the low-energy string effective
action, which can be written in $D$ dimensions
$$
S=-{1\over 16\pi G_D}\int d^D x\sqrt{|g|} e^{-\phi}(R+\pa_\mu\phi \pa^\mu
\phi -{H^2\over 12} +V) \eqno(2.13)
$$
(here $H=dB$ is the antisymmetric tensor field strength and $V$ is a
constant which is vanishing in a critical number
 of space-time dimensions).

The cosmological equations obtained from this action
for a homogeneous and spatially flat
background are characterized by a symmetry called "scale factor duality"
[10]. According to this symmetry, for any given set of solutions
$\{a_i(t), \phi(t), 1=1,...,D-1\}$ ($a_i$ are the scale factors of a
diagonal, not necessarily isotropic metric in the synchronous frame gauge
$g_{00}=1, g_{0i}=0$), the configuration obtained through the
transformation (for each $i$)
$$
a_i\ra \tilde a_i=a_i^{-1}~~~~,~~~~
\phi \ra \tilde \phi = \phi- 2 \ln a_i \eqno(2.14)
$$
is still a  solution of the graviton-dilaton system of equations
[10--12].

This transformation is just a particular case of a more general global
$O(d,d)$ covariance of the theory [13,14] ($d$ is the number of
coordinates upon which the background fields are explicitly independent,
in our case the $D-1$ spatial coordinates). In the case of anisotropic
backgrounds with generally non-diagonal metrics, this covariance, besides
transforming the dilaton field as
$$
\phi \ra \phi -\ln|\det g_{ij}| \eqno(2.15)
$$
also mixes non-trivially the components of the metric and of the
antisymmetric tensor as follows:
$$
M \ra \Om^T M\Om , \; \eqno(2.16)
$$
where $\Om$ represents a global $O(d,d)$ transformation, and
$$
M=\pmatrix{G^{-1} & -G^{-1}B \cr
BG^{-1} & G-BG^{-1}B \cr}
\eqno(2.17)
$$
($G \equiv g_{ij}$ and $B\equiv B_{ij}=-B_{ji}$ are matrix
representations of the $d$ by $d$
spatial part of the metric and of the
antisymmetric tensor, in the basis in which the $O(d,
d)$ metric is in off-diagonal form [6,13]).

In the case of manifolds with spatial sections of finite
volume (such as a torus), i.e. $(\int d^dx\sqrt{|G|})_{t=t_1}
=const<+\infty$, this  $O(d,d)$ covariance can be
 preserved even if the scalar poyential $V$ is taken dilaton-dependent,
 provided $\phi$ appears in the potential only
through the combination [13]
$$
\fb = \phi - \ln \sqrt{|G|} \eqno(2.18)
$$
(we have absorbed into $\phi$ the constant shift $-\ln \int d^dx$
required to secure the $GL(d)$ coordinate invariance of the corresponding
non-local action).

Moreover, the $O(d,d)$
covariance holds even if the equations are supplemented by
phenomenological source terms corresponding to bulk string matter [6].
Indeed, their effective equation of state can be expressed in terms of
the $2d$-dimensional phase-space variables $Z^A=(P_i, \pa X^i/\pa \sg)$,
  representing a solution of the string equations of motion: when the
background is changed according to eq. (2.16), the new solution $\tilde Z
= \Om^{-1}Z$ generates a new effective equation of state which leaves
the overall $O(d,d)$ symmetry unbroken [6].

The importance of scale factor duality, in our context, is that it
allows the construction of a mapping relating any decreasing curvature
background to a corresponding scenario in which the curvature is
increasing.

It is true, indeed, that the Hubble parameter is odd under
scale factor duality, $H\ra -H$, so that a duality transformation
$a\ra a^{-1}$ maps an expanding solution of Table I into a contracting
one of Table II (and vice versa). However,
the cosmological solutions with
monotonic evolution of the curvature (see for instance Section 3)
are in general defined on a half-line in $t$, namely $-\infty \leq t\leq
t_c$ and $t_c\leq t\leq \infty$, where $t_c$ is some finite value of the
cosmic time where a curvature singularity occurs [10]. We are thus
allowed, in such a case, to change simultaneously the sign of $\dot H$
and $\ddot a$ inside a given table (i.e. keeping the sign of $\dot a$
constant), by performing a duality transformation and by changing,
simultaneously, the domain of $t$ from $[t_c,\infty]$ to $[-\infty,
-t_c]$, i.e. by performing the inversion $t \ra -t$.

This means  that, to any given decelerated solution, with
monotonically decreasing curvature, typical of the standard
post-big-bang cosmology,
it is always possible to associate a pre-big-bang
solution, with accelerated evolution and increasing curvature, through
the transformation
$$
a(t) \ra a^{-1}(-t). \; \eqno(2.19)
$$
It is important to stress that this mapping property (valid  in the
cases of both expanding and contracting backgrounds) cannot be
 achieved in the
framework of Einstein's equations, where there is no dilaton.
 In that case,
  scale factor duality is broken; one is just left
 with the more conventional
time reversal symmetry $a(t) \ra a(-t)$.

Unlike straight scale-factor-duality,
the transformation (2.19) has non-trivial
fixed points $a(t)= a^{-1}(-t)$ describing  models that
 are monotonically
expanding (or contracting), since $H(t)=H(-t)$, and in which the
duration of the primordial superinflationary evolution equals that of
the dual decelerated phase, as $\dot H(t)=-\dot H(-t)$. We shall come
back on this point in the following section.
\vskip 1.5 cm

{\bf 3. An example of pre-big-bang dynamics}

Consider a spatially flat background configuration, with vanishing
antisymmetric tensor and dilaton potential. The dilaton field depends
only upon time, and the metric, which is assumed to describe a phase of
dimensional decoupling in which $d$ spatial dimensions expand  with
scale factor $a(t)$, and $n$ dimensions contract  with scale factor
$b(t)$, is given by
$$
g_{\mu\nu}= diag (1, -a^2(t)\da_{ij}, -b^2(t)\da_{ab}) \eqno(3.1)
$$
(conventions: $\mu,\nu=1,...,D=d+n$; $i,j=1,...,d$; $a,b=1,...,n$).

We are working in a synchronous frame in which $g_{00}=1$, $g_{0i}=0=
g_{0a}$, so that the time parameter $t$ coincides with the usual
cosmic time coordinate. In this gauge, the stress tensor of a comoving
source, in the perfect fluid approximation, becomes
$$
T_\mu\,^\nu =diag (\r(t),-p(t)\da_i^j,-q(t)\da_a^b), \; \eqno(3.2)
$$
where $p$ and $q$ are the pressures in the expanding and contracting
space, respectively. It is also convenient to introduce
the $O(d+n,d+n)$-invariant expressions for the dilaton
and the matter energy density,
that in this background take the form
$$
\fb = \phi - \ln \sqrt{|g|}=\phi -d \ln a -n  \ln  b \eqno(3.3)
$$
$$
\rb= \r \sqrt{|g|}= \r a^db^n \eqno(3.4)
$$
(we also define, for the pressure, $\pb =p\sqrt{|g|}$).

With these definitions, the equations obtained by varying the
effective action (2.13) (with $V=0$) and the action for the matter
sources,
$$
2(R_\mu\,^\nu +\big_\mu \big^\nu \phi)-{1\over 2}H_{\mu\a \b}H^{\nu\a
\b} = 16\pi G_D e^\phi T_\mu\,^\nu \eqno(3.5)
$$
$$
R-(\big_\mu \phi)^2+2\big_\mu \big^\mu \phi -{1\over 12}H_{\mu\nu\a}
H^{\mu\nu\a} =0
\eqno(3.6)
$$
$$
\pa_\nu(\sqrt{|g|}e^{-\phi}H^{\nu\a\b})=0\eqno(3.7)
$$
($\big_\mu$ is the Riemann covariant derivative), can be written in a
particularly simple form (when $B=0$).
The dilaton equation (3.6) reduces to
$$
\dot {\fb}^2 -2\ddot {\fb}+ dH^2+nF^2=0, \eqno(3.8)
$$
where $H=\dot a/a$, $F=\dot b/b$. Moreover, by combining this equation
with the time component of (3.5) we get
$$
\dot {\fb}^2-dH^2-nF^2=16\pi G_D\rb e^{\fb}. \eqno(3.9)
$$
{}From the space components of (3.5) we have finally the equations for the
external and internal pressure
$$
2(\dot H -H\dot {\fb})=16\pi G_D \pb e^{\fb} \eqno(3.10)
$$
$$
2(\dot F -F\dot {\fb})=16\pi G_D \overline q e^{\fb}. \eqno(3.11)
$$
[Note that the four equations (3.8) to (3.11) are
only a particular case of
the $O(d,d)$-covariant string cosmology equations of Ref. [6], for
vanishing $B$,$V$ and a perfect fluid description of the matter sources].

Suppose now that $a(t)$ describes accelerated expansion, $b(t)$
accelerated contraction, and that
$$
p=\ga_1 \r ~~~~,~~~~q=\ga_2 \r ~~~~,~~~~b=a^{-\ep} \eqno(3.12)
$$
with $\ga_1 , \ga_2 , \ep $ constants, $\ep >0$. "Stringy" sources,
asymptotically consistent with such background [5], must satisfy the
equation of state (2.12), which implies
$$
d\ga_1-n\ga_2=-1. \eqno(3.13)
$$

As discussed in [5], the three possible asymptotic behaviours,
$|q|<<|p|$,$|q|>>|p|$ and $|q|\sim |p|$ correspond to $
\ep <1$, $\ep >1$ and $\ep=1$, respectively. In the first two cases,
 however, eqs.
(3.8) to (3.11)  can be consistently solved only
 for $\r =p=q=0$, so that one
recovers the vacuum background [10] with a Kasner-like metric solution,
and $\fb \sim - \ln|t|$.

If we look for  solutions with non-vanishing bulk string matter we have
thus to consider the third possibility,  $b=a^{-1}$. In such a case
  eqs. (3.10) and (3.11) imply $p=-q$, and the condition (3.13) is
non-trivially satisfied by
$$
\ga_1=-\ga_2={-1\over d+n}.  \eqno(3.14)
$$
We can thus obtain, from eqs. (3.8) to (3.10),
the particular exact solution
$$
\eqalign{a(t)&= b^{-1}(t) = (-{t\over t_0})^{-2/(d+n+1)}, \cr
\phi =\phi_0& +2d~ \ln~a ~~~~~,~~~~~\r=\r_0 a^{n+1-d}, \cr}\eqno(3.15)
$$
where $t_0,\r_0,\phi_0$ are integration constants, related by
$$
\r_0 e^{\phi_0} = 4 {(d+n)(d+n-1)\over (d+n+1)^2}. \eqno(3.16)
$$

This solution satisfies the properties
$$
\eqalign{\dot a&>0~~~~,~~~~\ddot a>0~~~~,~~~~\dot H >0 \cr
\dot b&<0~~~~,~~~~\ddot b<0~~~~,~~~~\dot F <0 \cr}\eqno(3.17)
$$
Therefore, in the $t \ra 0_-$ limit, it describes a background in which
the evolution is accelerated, since $\{\dot a , \ddot a\}$ and
$\{\dot b ,
\ddot b\}$ have the same sign, and the curvature is growing, since
$\{\dot a , \dot H\}$ and $\{\dot b , \dot F \}$ have the same sign
(recall Tables I and II). Moreover, the equation of state of the sources,
$$
p=-q=-{\r\over d+n}\eqno(3.18)
$$
is compatible with the solution of the string equations of motion, so
this background is a typical example of {\it string-driven pre-big-bang}
cosmology.

It may be interesting to note that, in the four-dimensional isotropic
limit ($n=0,d=3$), the solution (3.15) simply reduces to
$$
a=({-t\over t_0})^{-1/2}~~~ ,~~~ \phi =\phi _0 -3 \ln(-{t\over t_0})~~~ ,
{}~~~\r= -3p=\r_0(-{t\over t_0}),
\eqno(3.19)
$$
namely to the superinflationary background obtained,
via
the duality transformation (2.14), from the standard
radiation-dominated cosmology (with constant dilaton),
$$
a=({t\over t_0})^{1/2}~~~~ ,~~~~ \phi =\phi _0 ~~~~ ,~~~~
\r= 3p=\r_0({t\over t_0})^{-2},
\eqno(3.20)
$$
as already pointed out in [6].

The simple example (3.15) can be used to illustrate two important
properties of the pre-big-bang phase: not only the curvature, but also
the total effective energy density and the temperature are growing
together with the curvature.

For the behaviour of $\r$ we have indeed, from eq. (3.15):
$$
\r(t)=\r_0(-{t\over t_0})^{2(d-n-1)/(d+n+1)}. \eqno(3.21)
$$
Since we are working in a Brans-Dicke frame, we have to include also the
time variation of the gravitational constant, $G\sim e^\phi$, in the
term that plays the role of the total gravitational source. We thus
obtain, for the effective energy density,
$$
G\r \sim \r_0 a^{d+n+1} = \r_0(-{t\over t_0})^{-2}, \eqno(3.22)
$$
which is always growing for $t \ra 0_-$ .

As far as the temperature behaviour is concerned,
it must be recalled
that, according to the string cosmology equations (3.5) to (3.7),
 the dilaton
gives no contribution to the covariant conservation
of the energy of the sources (in
spite of its coupling to the tensor $H_{\mu\nu\a}$ [6]).
Moreover, when $H_{\mu\nu\a}=0$, such a
conservation equation defines a regime of
adiabatic evolution for the perfect fluid.
 By using the standard thermodynamical arguments
(see for instance [21]) one can thus obtain, for any given equation of
state $p=\ga \r$, a relation between temperature and scale factor which,
in the case of $d$ isotropic spatial dimensions, takes the form
$$
a(T) \sim T^{-1/d\ga}. \eqno(3.23)
$$

The conservation equation, on the other hand, provides also a
 relation between scale factor and energy density:
$$
\r(a) \sim a^{-d(1+\ga)}. \eqno(3.24)
$$
For a standard radiation-dominated background ($\ga =1/d$), such as
 that of eq. (3.2), one then finds the usual
adiabatic decreasing of the temperature:
$$
\r \sim {1\over a^{d+1}} \sim T^{d+1}~~~~,~~~~ T\sim {1\over a}.
\eqno(3.25)
$$
For the dual background (3.19), on the contrary, $\ga =-1/d$ and the
temperature increases together with the scale factor,
$$
\r \sim {1\over a^{d-1}}\sim {1\over T^{d-1}}~~~~,~~~~T \sim a .
\eqno(3.26)
$$

The same result holds in the more general anisotropic pre-big-bang
background of eq. (3.15), with equation of state (3.18). Indeed in
 such a case
the conservation equation reads
$$
\dot {\rb} = H\rb \eqno(3.27)
$$
and implies
$$
\r(a) \sim a^{n+1-d}. \eqno(3.28)
$$
In terms of the total proper volume $V=V_dV_n=a^{d-n}$,
 eq. (3.27) can be
rewritten as an adiabaticity condition
$$
dS \equiv {1\over T}d(\r V)+{1\over T}(p{dV_d \over V_d}
+q{dV_n\over V_n})V
= {1\over T}d(\r V) + {p'\over T}dV=0 \eqno(3.29)
$$
corresponding to the effective pressure
$$
p'=p({d+n\over d-n})={\r \over n-d}. \eqno(3.30)
$$
The integrability condition $\pa^2S/\pa V\pa T = \pa^2 S/\pa T\pa V $
 then gives
$$
\r (t) \sim T^{n+1-d}, \eqno(3.31)
$$
which, using eq.(3.28), leads to
$$
T\sim a\sim (-t)^{-2/(n+1+d)}. \eqno(3.32)
$$
Thus we find that, during
the pre-big-bang phase, the temperature, also in the
anisotropic case,
 grows proportionally to the expanding scale factor.

Two comments are in order:

The first is that, according to this example, the temperature seems to
remain unchanged ($T=\ti T$) when passing from a background $a$ to the
dual one $\ti a = a^{-1}$. Indeed, from eqs. (3.25) and (3.26),
$$
\ti T \sim \ti a = a^{-1} \sim T. \eqno(3.33)
$$
This behaviour is to be contrasted with recent results, obtained in a
black-hole background, which suggest that the temperature should be
inverted under a duality transformation [22].

We must note, however, that the pre-big-bang solution considered here is
related to the decelerated post-big-bang phase not only through a
duality transformation, but also through a change of the cosmic time
range from $[-\infty,0]$ to $[0,\infty]$. Moreover, what we have
considered is the temperature of the sources, and not the horizon
temperature as in the black hole case.

Indeed, the appearance of a dynamical event horizon during
the pre-big-bang phase is not sufficient to introduce a thermal
 bath of geometric
origin, and to define an intrinsic "background temperature". The reason
is that, in contrast with the Rindler or de Sitter case, the Green
functions for fields embedded in a manifold with shrinking horizons are
not, in general, periodic with respect to the Euclidean time coordinate.
In other words, a De Witt detector [23] in the frame of a comoving
pre-big-bang observer will measure a particle background, but the
response function of the
detector is not of the thermal type and cannot define an intrinsic
temperature.

Moreover, even
introducing (na\"\i vely) a temperature associated with the pre-big-bang
event horizon in terms of its local surface gravity, $T_H\sim
|t|^{-1}$, it would be impossible in this cosmological context
to discuss the transformation of the
temperature under duality by using the
horizon thermodynamics, because there is no event horizon in the dual,
decelerated phase.

It is possible, however, to obtain a thermal spectrum from the geometry,
in a cosmological context, by considering the high frequency sector of
the particle spectrum produced by a background that evolves smoothly
between two asymptotically flat states [23--25]. In this way one can
associate a temperature with the geometry, independently from the presence
of event horizons, and one can easily
show that for a "self-dual" metric [i.e.
a metric that satisfies $a^{-1}(t)=a(-t)$, see for instance Section. 5],
this "geometric" temperature is inverted under duality, just as in the
black hole case.

The demonstration is based on the properties of the Bogoliubov
coefficients, $c_+(k)$ and $c_-(k)$, parametrizing for each mode
$k$ the transformation between the  $|in>$ and $|out>$ vacuum,
and normalized in such a way that
 $|c_-(k)|^2=<n_k>$ measures the
spectral number of produced particles, and $|c_+|^2=1+|c_-|^2$.

Consider indeed the transformation $t \ra -t$, connecting an
expanding to a contracting background. The
$|in>$ and $|out>$ states get interchanged, but
the constant equilibrium temperature $T_0$ associated with the
thermal distribution of the produced particles is preserved,
as it depends on the norm of the Bogoliubov coefficients, which
is left unchanged
$$
|c_+|=|<in,+|out,+>|~~~~,~~~~|c_-|=|<in,+|out,->|.  \eqno(3.34)
$$

The local temperature $T(t)=T_0/a(t)$,
characteristic of the given geometry, changes however from a
regime of adiabatic red-shift to an adiabatic blue-shift,
$T(t) \ra
T(-t)=T_0/a(-t)$. If the metric is
self-dual we   obtain, in units of $T_0$,
$$
T(a(-t))=T(a^{-1}(t))\equiv \ti T(t) = a(t)={1\over T(a(t))}\eqno(3.35)
$$
so that the temperature and its dual $\ti T$ satisfy
$$
T\ti T = T^2_0 = const,  \eqno(3.36)
$$
in agreement with earlier arguments of
string thermodynamics in a cosmological context [2,3].

The second comment regards the growth of the temperature of the source
during the pre-big-bang phase [see eqs. (3.25) and (3.26)]. In the
regime of
radiation-dominated expansion, $aT$ stays constant, and
the usual way to get
the very large present value of the entropy, starting from reasonable
initial conditions, is through the occurrence of
some non-adiabatic "reheating" era.
In the dual pre-big-bang regime, on the
contrary, the temperature of the sources grows with the scale factor
just because of the adiabatic evolution. It is thus possible (at least
in principle) for the Universe to emerge from the big-bang with
 hot enough sources  and large enough scale factor  to solve the
entropy and the horizon problems.

Consider, for example, a cosmological evolution which is symmetric
around the phase of maximal curvature, in the sense that
 the post-big-bang decelerated expansion
of the three-dimensional space  [eq. (3.20)]
is preceeded by a corresponding phase of
 accelerated expansion [eq. (3.19)], which extends in time
 as long as the dual decelerated phase.

The pre-big-bang superinflation ends at the maximum curvature scale,
which is expected to be of the order of the Planck mass, $H_f\sim M_P$.
On the other hand, our present curvature scale $H_0$ is about $60$
orders of magnitude below the Planck scale. In the case of a symmetric
evolution evolution around the maximum (see e.g. the example
presented in Section 5), the pre-big-bang superinflation era
must start at a
scale $H_i \me H_0 \sim 10^{-60}M_P$, thus providing a total amount of
inflation
$$
{a_f\over a _i}\sim ({H_f\over H_i})^{1/2} \Me 10^{30}, \eqno(3.37)
$$
which is just the one required for the solution of the standard
cosmological puzzles [26].

Apart from this amusing numerical coincidence, what is important to
stress is that in the case of a symmetric temporal
 extension the pre-big-bang scenario seems capable
 of providing automatically the required
amount of inflation, {\it for any given value of the final
observation time $t_0$}, thus avoiding the anthropic problem that
 is in
general included in the more conventional, post-big-bang, inflationary
models (see [27] for a lucid discussion of the anthropic principle in an
inflationary context).

It should be clear, however, that even in this context the large
present value of the total entropy would be the consequence of a
non-adiabatic conversion of the hot string gas into hot radiation,
occurring during the transition to the standard scenario. A
possible example of such process is the decay of highly excited
string states, created by the background evolution through
the mechanism described (for the case of gravitons) in the following
section. One should also mention, in this context, the existence of an
additional source  of radiation entropy due to the presence of
non-trivial antisymmetric tensor backgrounds. Indeed, even if the
evolution is globally adiabatic, entropy exchanges may occur between
$B$ and the fluid part of the source, because of the covariant
conservation of the total energy density. As a consequence, the
damping of $B$ in time (which is expected to occur if the standard
isotropic scenario is to be recovered) should be accompanied
by an entropy increased in the radiation fluid, as already
stressed in [6].

\vskip 1.5 cm

{\bf 4. Observable tracks of a pre-big-bang phase}

As discussed in the previous sections, there are various
reasons for which the occurrence of a period of accelerated evolution
and growing curvature, in the past history of our Universe, may seem
both plausible (from a string theory point of view) and attractive
(for its phenomenological aspects). The important question to ask,
therefore, is whether the occurrence of such a phase may be tested
experimentally in some way.

An affirmative answer to this question is provided by the study of
the cosmic graviton background [28--30]: indeed, gravitons decoupled from
matter earlier than any other field (nearly at the Planck scale), so
that the shape of the spectrum of the gravitons produced by the
transition from the pre-big-bang to the post-big-bang scenario should
survive, nearly unchanged, up to the present time. Given a model of
cosmological evolution, in particular, one can compute the expected
graviton spectrum, and then constrain the model by using the present
observational bounds.

Consider, for example, a generic model of pre-big-bang evolution in
which the dilaton is growing, $d$ dimensions expand with scale factor
$a(\eta)$ and $n$ dimensions contract with scale factor $b(\eta)$ in
such a way that, for $\eta<-\eta_1<0$,
$$
a\sim (-\eta)^{-\a}~~~~,~~~~b\sim (-\eta)^{\b}~~~~,~~~~\phi \sim
\ga \ln a ,  \eqno(4.1)
$$
where $\eta$ is the conformal time coordinate, related to the cosmic
time $t$ by $dt/d\eta=a$ (we shall denote, in this section, derivatives
with respect to $\eta$ with a prime). Note that in eq. (4.1)
$\eta$ ranges over negative values, so that $\a ,\b$ and $\ga$ are all
positive parameters. We shall assume that this phase is followed (at
$\eta= - \eta_1$) by the standard radiation-dominated,
 and matter-dominated
(at $\eta=\eta_2$), expansion of three spatial dimensions, with frozen
dilaton and radius of the internal space, namely
$$
\eqalign{a&\sim \eta~~~~,~~~~b=1~~~~,~~~~\phi=\phi_0~~~~,~~~~
-\eta_1<\eta<\eta_2 , \cr
a&\sim \eta^2~~~~,~~~~b=1~~~~,~~~~\phi=\phi_0~~~~,~~~~\eta>\eta_2>0.
\cr}\eqno(4.2)
$$

The spectrum of the gravitons produced from the vacuum, because of the
background variation, is to be computed from the free propagation
equation of a metric fluctuation, $h_{\mu\nu}=\da g_{\mu\nu}$, obtained
by perturbing the background equations at fixed sources [31--33] (in our
case dilaton and antisymmetric tensor included),
$\da T_\mu^\nu =\da \phi=\da H_{\mu\nu\a}=0$. The contribution to the
graviton production from a possible variation of the effective
gravitational coupling can be accounted for, in general, by perturbing a
Brans-Dicke background [34]. In our case we shall perturb the string
gravity equations (3.5) to (3.7) around the cosmological configuration
$\{H_{\mu\nu\a}=0, \phi=\phi(t), g_{\mu\nu}=diag(1,g_{ij},g_{ab})\}$,
where $g_{ij}=-a^2\da_{ij}$ and $g_{ab}=-b^2\da_{ab}$ are the $d$- and
$n$-dimensional conformally flat metrics of the expanding and
contracting manifolds, respectively.

By imposing the gauge
$$
h_{0\mu}=0~~~~,~~~~\big _\nu h_\mu\,^\nu =0~~~~,~~~~g^{\mu\nu}h_{\mu\nu}
=0, \eqno(4.3)
$$
and by considering a perturbation $h_i\,^j(\vec x,t)$ propagating in the
expanding "external" space, one then gets the linearized equation [34]
$$
g^{\mu\nu}\big_\mu\big_\nu h_i\,^j -\dot \phi \dot h_i\,^j=0, \eqno(4.4)
$$
where $\big_\mu$ denotes the covariant derivative with respect to the
background metric $g$. By performing a Fourier expansion, and
introducing the variable
$$
\psi_i\,^j=h_i\,^j b^{n/2} a^{(d-1)/2} e^{-\phi/2},  \eqno(4.5)
$$
one can conveniently rewrite eq. (4.4), for each mode $\psi_i\,^j(k)$
in terms of the conformal time variable, in the form:
$$
\psi^{\se}(k)+[k^2-V(\eta)]\psi(k)=0, \eqno(4.6)
$$
where
$$
\eqalign{
V(\eta)&= {d-2 \over 2}{a^{\se}\over a}+{n\over 2}{b^{\se}\over b}
-{\phi^{\se}\over 2} +{1\over 4}(d-1)(d-3)({a^\pr \over a})^2
+{n\over 4}(n-2)({b^\pr \over b})^2 \cr
&+ {1\over 4}\phi^{\pr 2}
+{n\over 2}(d-1){a^\pr \over a}{b^\pr \over b}
-{1\over 2}(d-1){a^\pr \over a}\phi^\pr -{n\over 2}{b^\pr \over b}
\phi^\pr \cr}
\eqno(4.7)
$$

For the phenomenological background (4.1), (4.2) we thus have  the
effective potential barrier
$$
\eqalign{
V(\eta)={1\over 4\eta^2}\{[\a (d-1-\ga)-n\b +1]^2-1\}~~~~,~~~~
\eta&<-\eta_1 \cr
V(\eta)=0~~~~~~~~~~~~~~~~~~~~~~~~~~~~~~,~~~~~~~-\eta_1<\eta&<\eta_2 \cr
V(\eta)={2\over \eta^2}~~~~~~~~~~~~~~~~~~~,~~~~~~~
{}~~~~~~~~~~~\eta&>\eta_2,  \cr}
\eqno(4.8)
$$
which clearly displays the usual contribution of the accelerated
expansion ($\a \not= 0$) [28,29], of the dimensional reduction process
($n\b \not= 0$) [35,36] and, in addition, of the dilaton variation
($\ga \not=0$) [34].

The general solution of eq. (4.6) can be written in terms of the first-
and second-kind Hankel functions $H_\nu^{(1)}(k\eta), H_\nu^{(2)}(k\eta)
$, of index
$$
\nu ={1\over 2}[\a (d-1-\ga)-n\b +1]\eqno(4.9)
$$
during the pre-big-bang phase, and of index $|\nu|=1/2$ and $|\nu|=3/2$
during the radiation- and matter-dominated phase, respectively. Starting
with the initial solution
$$
\psi(k)=N\eta^{1/2}H_\nu^{(2)}(k\eta)~~~~,~~~~\eta<-\eta_1 \eqno(4.10)
$$
representing positive frequency modes in the $\eta \ra -\infty$ limit
(and corresponding to the Bunch-Davies "conformal" vacuum [23]), one has
in general a linear combination of positive ($H^{(2)}$) and negative
($H^{(1)}$) frequency solutions for $\eta \ra \infty$. The Bogoliubov
coefficients $c_{\pm}(k)$, describing such mixing, can be obtained by
matching the solutions at $\eta=\eta_1$ and $\eta=\eta_2$ ($\eta_2>>
\eta_1$).

For $k\eta_1<1$, using the small
argument limit of the Hankel functions, and considering only the first
background transition, this "sudden" approximation gives
$$
|c_-(k)|\simeq (k\eta_1)^{-|\nu+1/2|} \eqno(4.11)
$$
(we neglect numerical factors of order unity). For $k\eta_1\Me1$, i.e.
for modes with a comoving frequency $k$ higher than  the
potential barrier $[V(\eta_1)]^{1/2}\simeq \eta_1^{-1}$, this
approximation is no longer valid; then the mixing coefficient $c_-$ is to
be computed by replacing the potential step with a smooth transition
from $V(\eta_1)$ to $0$ (in order to avoid ultraviolet divergences).
In this way one finds, however, that for $k\eta_1>1$ the mixing is
exponentially suppressed [23,25,35], so that it may be neglected for our
purposes.

The modes with  sufficiently small frequency $k<1/\eta_2$ are
significantly affected also by the second background transition, from
the radiation-dominated to the matter-dominated regime [28,37-39].
In the same approximation one finds that, for this frequency sector,
$$
|c_-(k)|\simeq (k\eta_1)^{-|\nu+1/2|}(k\eta_2)^{\mp1},  \eqno(4.12)
$$
where the sign of the second exponent is $-1$ ($+1$) if $\nu+1/2>0$
($<0$) [34]. For all the models of dynamical dimensional decoupling
 proposed  up
to now (see for instance [5,15,40]), including the pre-big-bang
example of the previous section, one has however $\nu+1/2>0$. We shall
thus disregard here, for simplicity, the alternative possibility (but we
note that for $\nu+1/2<0$ the phenomenological constraints on the graviton
spectrum that we shall present below turn out to be somewhat relaxed
 [34]).

The spectral energy density $\r(\om)=\om(d\r_g/d\om)$, which is the
variable usually adopted to characterize the energy distribution of the
produced gravitons [28,29,39], is simply related to the Bogoliubov
coefficient $c_-(k)$ that gives, for each mode, the final number of
produced particles:
$$
\r(\om)\simeq\om^4|c_-|^2 , \eqno(4.13)
$$
where $\om=k/a(t)$ is the proper frequency (again we neglect numerical
factors of order unity). On the other hand, $|\eta|^{-1}\simeq
|a(\eta)H(\eta)|$; moreover, $\eta_1$ corresponds to the beginning of
the radiation era, so we can conveniently express the maximum curvature
scale $H_1\equiv H(\eta_1)$, reached at the end of the pre-big-bang
phase, as $H_1^2\simeq G\r_\ga(\eta_1)$, where $G$ is the usual Newton
constant, and $\r_\ga$ is the radiation energy density [29]. It follows
that the spectral energy density (4.13), at the present observation time
$t_0$, and in units of critical energy density $\r_c$, can be finally
expressed as
$$
\eqalign{
\Om(\om,t_0)&\equiv {\r(\om)\over \r_c}\simeq GH_1^2\Om_\ga(t_0)
({\om\over \om_1})^{4-2|\nu+1/2|}~~~,~~~~~~~~~\om_2<\om<\om_1 \cr
\Om(\om,t_0)&\equiv {\r(\om)\over \r_c}\simeq GH_1^2\Om_\ga(t_0)
({\om\over \om_1})^{4-2|\nu+1/2|}({\om\over \om_2})^{-2}
{}~~~,~~~\om_0<\om<\om_2 \cr}\eqno(4.14)
$$
Here $\Om_\ga(t_0)\sim 10^{-4}$ is the fraction of critical energy
density present today in radiation form; $\om_0\sim H_0\sim10^{-18}$
Hz is the minimal frequency inside the present Hubble radius;
$$
\om_2={H_2a_2\over H_0 a_0}\om_0 \sim 10^2 \om_0
\eqno(4.15)
$$
is the frequency corresponding to the radiation $\ra$ matter transition
and, finally,
$$
\om_1={H_1a_1\over H_0a_0}\om_0\sim10^{29}({H_1\over M_P})^{1/2}\om_0
\eqno(4.16)
$$
is the maximum cut-off frequency depending on the height of the
potential barrier $V(\eta_1)$, i.e. on the big-bang curvature scale
$H_1$.

The two important parameters of the spectrum (4.14) are the maximum scale
$H_1$  and the power $\nu$ which depends on the pre-big-bang kinematics
and fixes the frequency behaviour of the energy density of the relic
gravitons. In four dimensions ($d=3, n=0$), and in the absence of
dilaton contributions ($\ga=0$), the high-frequency behaviour of the
spectrum mimics exactly the behaviour of the curvature scale for $\eta<-
\eta_1$, as was stressed in [29,39] (we recall indeed that in such a case
$4-2|\nu+1/2|=2-2\a$, so that a  de Sitter phase at constant curvature,
$\a=1$, corresponds to a flat spectrum). If, however, the other
contributions to the graviton production are included, then the total
spectrum may be flat or decreasing even if the curvature is growing.

What is important to stress is that for a flat or decreasing spectrum the
most significant bound is provided by the isotropy of the
electromagnetic background radiation [28], $\Om<\Om_i$, while if the
spectrum is growing fast enough it is only constrained by the critical
density bound [34,39], $\Om<\Om_c$. The first bound turns out to be most
constraining if imposed at the minimum frequency $\om_0$, the second
bound at the maximum frequency $\om_1$ (there is also a bound obtained
from the pulsar-timing data [28], to be imposed at a frequency $\om\sim
10^{-8}$ Hz, but it is not significant in our context [34]).

The two constraints
$$
\Om(\om_0)<\Om_i~~~~~,~~~~~\Om(\om_1)<\Om_c \eqno(4.17)
$$
define an allowed region in the $(H_1,\nu)$ plane, which is bounded by
the curves [obtained from eq.(4.14)]
$$
\eqalign{
Log({H_1\over M_P})&<2+{1\over 2} Log\Om_c , \cr
Log({H_1\over M_P})&<{1\over |\nu+1/2|}(116+Log\Om_i)-58 . \cr}\eqno(4.18)
$$
The region allowed by the present experimental data ($\Om_c\sim 1,
\Om_i\sim 10^{-8}$)  is shown in {\bf Fig. 2}.

Because of the many approximations made, and of the uncertainties in the
experimental data, this figure is expected to give, of course, only a
qualitative picture of the possible phenomenological scenario. We can
see, nevertheless, that there is a maximum allowed scale of curvature,
$H_1\sim 10^2M_p$, which can be reached only if the spectrum is growing
fast enough, $|\nu+1/2|\me 1.8$. The general bound on the pre-big-bang
kinematics to allow a given final scale $H_1\leq 10^2M_P$,
in particular, is given by eq. (4.18) as
$$
|\nu+{1\over 2}|\leq {108\over 58+Log (H_1/M_P)}. \eqno(4.19)
$$
The Planck scale, in particular, can be reached if $|\nu+1/2|\me 1.88$.
Flat or decreasing spectra, $|\nu+1/2|\geq 2$, can only be extended to a
maximum scale $H_1\me 10^{-4}M_P$. One thus recovers the well-known
bound [41] (already quoted in Section 1) that applies to a
four-dimensional primordial
phase of the de Sitter type (which has $\nu+1/2=2
$, and thus corresponds   to a flat spectrum).

This bound is avoided by the
  pre-big-bang example given in Section .2. In that case
$$
\a=\b={2\over 3+d+n}~~~~~,~~~~~\ga=2d ,  \eqno(4.20)
$$
and one gets an increasing-with-energy graviton spectrum  with
$$
\nu+{1\over 2}={2\over 3+d+n}. \eqno(4.21)
$$
As a result, the curvature can reach the maximum allowed
 scale for any number
of dimensions.

The situation that we have discussed refers to the case of a
"sudden" transition from the growing curvature regime to the decreasing
one. The transition, however, could pass through an intermediate de
Sitter phase of maximal constant curvature. In such a case,
 the high-frequency part of the graviton spectrum is flattened,
 while the lower-frequency part continues to grow.
  It is still possible to evade the bound
$H_1\me 10^{-4}M_P$, and to consider a de Sitter inflation occurring
at the Planck scale, but its duration turns out to be constrained by the
present pulsar-timing data, in such a way that $10^{-1}M_P$ is
practically the
maximum  allowed value of $H_1$ in such a Planck-scale inflation [39].

Note that, in this paper, we are using the data on the
isotropy of the electromagnetic
 CMB only as a bound on the possible
presence of a relic graviton background, since we are mainly interested
in the limiting value of the curvature for a phase of pre-big-bang
evolution. It is also possible, however, to interpret the CMB anisotropy
recently detected by COBE [42] as entirely (or at least partially, but
significantly) due to a stochastic background of cosmic gravitons
[43,34]. By making such an assumption, and by using the graviton
spectrum corresponding to our pre-big-bang model (4.21), the COBE
result can be read as providing an interesting relation between the
maximum scale $H_1$ and the total number of spatial dimensions
$d+n$.

We should mention, finally, that an additional phenomenological
signature of a pre-big-bang phase is contained in the squeezing
parameter $r(\om)$ which characterizes the quantum state of the
relic cosmic
gravitons [44], produced from the vacuum by the pre-big-bang
$\ra$ post-big-bang transition.

Such a parameter can be approximately expressed in terms of the
Bogoliubov coefficients as [34,44]
$$
r(\om)\simeq \ln|c_-| . \eqno(4.22)
$$
For the high-frequency part of the spectrum ($\om>\om_2$)  we thus  find,
from eq. (4.11),
$$
r(\om)\simeq |\nu+{1\over 2}|[25-\ln({\om\over Hz})+{1\over 2}
\ln({H_1\over M_P})]. \eqno(4.23)
$$
The first term on the right-hand side is expected to be dominant (at
least in the presently allowed frequency range for a possible graviton
detection); the second term determines the variation with frequency of
the squeezing, and the last term provides a correction if the final
curvature scale differs from the Planck one. In the case of a future
{\it direct} observation of the cosmic graviton background, a measure of
the parameter $r(\om)$ would provide information  both on the maximum
curvature scale $H_1$  and on the kinematics of a possible pre-big-bang
phase.
\vskip 1.5 cm

{\bf 5. The high curvature regime}

In the examples of pre-big-bang scenarios so far considered the growth
of the curvature, as well as  that of the
 effective coupling $e^\phi$, were
unbounded [see for instance eq. (3.15)]. According to the
phenomenological constraints discussed in the previous section, however,
this growth must stop, at the latest around the Planck scale. A realistic
model for this scenario should thus describe also a smooth transition
from the growing to the decreasing curvature phase, avoiding the
singularity both in the curvature and in the dilaton field.

Of course, in the high-curvature regime in which one approaches the
Hagedorn and Planck scale, string corrections (of order $\ap$ and higher)
are expected to modify the na\"\i ve low-energy effective action (2.13).
Moreover, in such regime a more realistic cosmology is probably obtained
by including the contributions of a dilaton potential $V(\phi)$, and
also of the antisymmetric tensor field. Indeed, there are regular
examples (without sources) in which both a constant dilaton potential
[12]  and an $O(d,d)$-generated torsion background [45] contribute to
violate the strong energy condition and to avoid the singularity.

Let us resort, therefore, to the full $O(d,d)$-covariant equations of
string cosmology, obtained from the action (2.13) (supplemented by a
phenomenological source term describing bulk string matter). Let us
assume, furthermore, that the $d$-dimensional space has finite volume
(even if flat, for simplicity) such as a torus. It is thus possible
to make  the  $O(d,d)$ covariance of the background
fields compatible with the $GL(d)$ coordinate invariance,
 even in the presence
of a non-trivial dilaton potential, provided
$V=V(\fb)$, where $\fb$ is as defined in eq. (2.18). In such case
the field equations, obtained from the total (generally non-local)
action, can be written in explicit $O(d,d)$-covariant form as [6]
$$
\dot{\fb}^2-2\ddot{\fb}-{1\over 8}Tr(\dot M \eta )^2+{\pa V
\over \pa \fb}-V=0\eqno(5.1)
$$
$$
\dot{\fb}^2+{1\over 8}Tr(\dot M \eta )^2-V=\rb e^{\fb}\eqno(5.2)
$$
$$
(e^{-\fb}M\eta \dot M)\dot{~} =\overline T \eqno(5.3)
$$
where $M$ is the matrix defined in eq. (2.17), $\overline T$ is a $2
d\times 2d$ matrix representing the spatial part of the string stress
tensor (including the possible antisymmetric current density, source of
torsion [6]), and $\eta$ is the $O(d,d)$ metric in off-diagonal form
$$
\eta =\pmatrix{0&I\cr I&0\cr}\eqno(5.4)
$$
(Note that, throughout this section, both $\overline T$ and $\overline
\r$ are expressed in units of $16\pi G_D$, so that they have
dimensions $L^{-2}$.)

These three equations generalize, respectively,
the dilaton equation (3.8), and the time and space part of the
graviton equations, (3.9) to (3.11),
to which they reduce exactly for $V=0$,
vanishing torsion, and diagonal metric background. Their combination
provides the useful covariant conservation equation of the source energy
density, which can be written in compact form as [6]
$$
\dot{\rb}+{1\over 4}Tr(\overline T \eta M\eta\dot M \eta)=0. \eqno(5.5)
$$

As in ordinary cosmology, this system of
equations must be supplemented by an equation of state, which
in this context reads as a relation of the form
$\overline T=\overline T(\rb ,M)$.
In order to provide an explicit
example of regular pre-big-bang scenarios, we note
that the variables of eqs. (5.1) to (5.3) can be separated,
 and the equations can be
solved, for any given $\overline T/ \rb$, provided
$$
{\pa V\over \pa \fb }=2V. \eqno(5.6)
$$

Indeed, using this condition, the combination of eqs. (5.1) and (5.2)
gives
$$
(e^{-\fb})\ddot{~}={1\over2} \rb .  \eqno(5.7)
$$
Defining a time coordinate $\xi$ such that
$$
d\xi={\rb \over \r_0}dt\eqno(5.8)
$$
($\r_0$ is a constant, and we shall denote with a prime differentiation
with respect to $\xi$), eqs. (5.7) and (5.3) can then be easily
integrated a first time to give
$$
\rb (e^{-\fb})^\pr ={1\over 2}\r_0^2(\xi +\xi_0)\eqno(5.9)
$$
$$
\rb M\eta M^\pr=\r_0^2\Ga e^{\fb}. \eqno(5.10)
$$
where
$$
\Ga =\int {\overline T\over \rb}d\xi \eqno(5.11)
$$
($\xi_0$ is an integration constant).

On the other hand, by using (5.10)
to eliminate $M\eta\dot M$, and by noting that $\overline T/\rb = \Ga^
\pr$, we can rewrite the conservation equation as
$$
{\rb}^{\pr} e^{-\fb} =-{1\over 8}\r_0^2Tr(\Ga\eta\Ga\eta)^\pr .
 \eqno(5.12)
$$
By adding eqs. (5.9) and (5.12), integrating a second time
and defining (for later
convenience)
$$
D(\xi)=4\b +(\xi+\xi_0)^2-{1\over 2}Tr(\Ga\eta)^2 \eqno(5.13)
$$
($\b$ is an integration constant), we obtain finally
$$
\rb e^{-\fb}={1\over 4}D\r_0^2 , \eqno(5.14)
$$
which, inserted into (5.9) and (5.10), gives
$$
\fb^{\pr}=-{2\over D}(\xi+\xi_0)\eqno(5.15)
$$
$$
M\eta M^\pr={4\Ga \over D}. \eqno(5.16)
$$

We have thus satisfied eq. (5.3), and a combination of eqs. (5.1) and
(5.2). We must still impose eq. (5.2) which, using eqs. (5.15) and (5.16)
 and
the identity [6]
$$
(M\eta M^\pr \eta)^2=-(M^\pr \eta)^2 \eqno(5.17)
$$
reduces to the condition
$$
(\xi +\xi_0)^2-{1\over 2}Tr(\Ga \eta)^2=D+{VD^2\over 4}({\r_0\over \rb})
^2 . \eqno(5.18)
$$

This condition, together with eq. (5.6), can be satisfied in two ways.
The first way is trivial, $V=0$ and $\b=0$. The equations (5.15)
and (5.16) can then be
integrated for any given $\Ga(\xi)$, but the singularity
in this case cannot be avoided, because of the zeros of $D(\xi)$.

The second possibility corresponds to a non-trivial two-loop potential
$V(\fb)=-V_0e^{2 \fb}<0$, with $\b=V_0$. In this case too eqs. (5.6)
and (5.18) are both
satisfied, and we can obtain examples of
backgrounds that are exact solutions of the string cosmology equations
(5.14)--(5.16) (with $\b=V_0$), and which describe a smooth evolution from
asymptotically growing to asymptotically decreasing curvature, without
singularities.
We shall give some examples below.

Let us consider, first of all, a $d$-dimensional isotropic
background, with scale factor $a$, vanishing torsion ($B=0$), and
sources with a diagonal stress-tensor, so that [6]
$$
M\eta M^\pr=2{a^\pr \over a}\pmatrix{0&I\cr -I&0 \cr}~~~,~~~
\overline T=\pb \pmatrix{0&I\cr -I&0\cr}, \eqno(5.19)
$$
where $I$ is the $d$-dimensional identity matrix. By defining
  $\pb /\rb =
\ga$ we   thus have
$$
D=4V_0+(\xi+\xi_0)^2-d(\int \ga d\xi)^2 . \eqno(5.20)
$$

The integration of the cosmological equations is thus
immediate in the case
of a perfect fluid source, with $\ga=const$. Consider for example the
interesting case of the radiation-like solution,
$$
\ga={1\over d}~~~~,~~~~\int \ga d\xi={1\over d}(\xi+\xi_1)~~~~,~~~~
\xi_1=const.  \; ,\eqno(5.21)
$$
and define, for convenience, the constant parameters
$$
\a ={d-1\over d}~~~,~~~b=2({\xi_0\over \xi_1}-{1\over d})~~~,~~~
c={4V_0+\xi_0^2\over \xi_1^2}-{1\over d} . \eqno (5.22)
$$
By choosing the arbitrary constants $\xi_0$ and $\xi_1$ in such a way
that
$$
\Da^2\equiv 4\a c -b^2>0\eqno(5.23)
$$
the integration of eqs. (5.15) and (5.16)  then provides  the non-singular
solution
$$
\fb =\phi_0+\ln(\a{\xi^2\over \xi_1^2}+b{\xi\over \xi_1}+c)^
{-{d\over d-1}}-{4(\xi_1-\xi_0)\over (d-1)\Da \xi_1}\arctan
({2\a \xi+b\xi_1\over \Da \xi_1}) \eqno(5.24)
$$
$$
a=a_0(\a{\xi^2\over \xi_1^2}+b{\xi\over \xi_1}+c)^
{{1\over d-1}}\exp\{ {4(\xi_1-\xi_0)\over (d-1)\Da \xi_1}\arctan
({2\a \xi+b\xi_1\over \Da \xi_1})\} \eqno(5.25)
$$
($a_0$ and $\phi_0$ are dimensionless integration constants). Their
combination gives
$$
e^\phi=
a_0^d e^{\phi_0}
\exp\{ {4(\xi_1-\xi_0)\over (d-1)\Da \xi_1}\arctan
({2\a \xi+b\xi_1\over \Da \xi_1})\},  \eqno(5.26)
$$
which is growing for $\xi_1>\xi_0$.

Note that the dilaton potential $V(\fb)$ provides its strongest
contribution just in correspondence of the phase of maximum
curvature, while it rapidly decays to zero away from this regime. We
 thushave  an example   of "running cosmological constant"
$\La  = V(\fb)$, which
may suggest a new possible approach to the cosmological constant
problem based on the damping  of $\La$ obtained from a
non-local effective action (see also [46]).

It is
also interesting to point out that, in the case of a radiation-like
source, the time coordinate $\xi$ coincides with the conformal time
$\eta$. From eq. (5.14) we  indeed obtain, for $\rb$
$$
\rb= {\xi_1^2\r_0^2 e^{\phi_0}\over
4(\a{\xi^2\over \xi_1^2}+b{\xi\over \xi_1}+c)^
{{1\over d-1}}}
\exp\{- {4(\xi_1-\xi_0)\over (d-1)\Da \xi_1}\arctan
({2\a \xi+b\xi_1\over \Da \xi_1})\}
, \eqno(5.27)
$$
so that, from the definition of $\xi$,
$$
{d\xi\over dt}={\rb \over \r_0}={1\over 4}\xi_1^2\r_0e^{\phi_0}({a_0
\over a}).  \eqno(5.28)
$$

Equations (5.25) and (5.26) describe a radiation-dominated solution,
which interpolates without singularities between an initial asymptotic
phase of {\it accelerated contraction}, $a\sim (-t)^{2/(d+1)}$,
$\phi=\phi_1=const.$, {\it growing curvature}, and a final phase of
{\it decelerated expansion}, $a\sim t^{2/(d+1)}$, $\phi=\phi_2=const.$,
{\it decreasing curvature}. There are no horizons, and both the
curvature and the dilaton field are regular.

In this example the final expanding state is reached starting from an
initial contraction. There is, however, also the dual solution
$\ti\phi , \ti a$ corresponding to the equation of state
$$
\pb = -{\rb\over d}~~~~,~~~~
\int\ga d\ti\xi=-{1\over d}(\ti\xi+\xi_1) , \eqno(5.29)
$$
where $\ti\xi$ is the "dual" time coordinate defined by eq. (5.8) for
this new solution. From eqs.(5.14)--(5.16) one obtains, in terms of
$\ti\xi$, that the new solution $\{\ti\phi,\ti a\}$ is related to the
old one $\{\phi,a\}$ by
$$
\eqalign{
\ti{\fb}&=\fb(\ti\xi) \cr
({\ti a\over a_0})&=a_0a^{-1}(\ti\xi) \cr
\ti{\rb}&={1\over 4}\xi_1^2\r_0^2e^{\phi_0}{a_0\over a(\ti\xi)}=
{1\over 4}\r_0^2e^{\phi_0}({\ti a\over a_0}) \cr}, \eqno(5.30)
$$
where $d\ti\xi \sim \ti a dt$. In this solution, therefore, the
background evolves smoothly from an initial accelerated phase of
{\it superinflationary expansion}, $a\sim (-t)^{-2/(d+1)}$, with a
logarithmically {\it increasing dilaton},
towards a final asymptotic phase of {\it
decelerated contraction}, $a\sim t^{-2/(d+1)}$, a logarithmically
{\it decreasing dilaton}, and {\it decreasing curvature}.
Again the maximum
curvature regime is crossed over without singularities.

 These two kinds of solution   smoothly  connect
the pre-big-bang phase to the post-big-bang one by changing the sign of
$H$ (from expansion to contraction, and vice-versa). Their existence
 may suggests a
possible scenario in which the transition to the standard cosmology
occurs after a period of background oscillations [11,47],
namely a series of
anisotropic contractions and expansions in which "external" and
"internal" coordinates exchange their roles  periodically.
Such a scenario may also recall the Kasner-like oscillating
behaviour, typical in general relativity of a cosmological metric
approaching the initial singularity [48], with the only difference that
in our case the singularity is smoothed out by a phase of maximal
curvature.

This possibility is suggested by solutions obtained for a
 fixed equation of
state. We might expect, however, that the transition from a given
kinematical class of background evolution to the dual one might be
associated with a corresponding transition between
  duality-related regimes also in the matter sources.
In such a case it becomes possible
to go across the phase of maximum curvature, without singularities, even
for monotonically
expanding (or contracting) "self-dual"
solutions.

Consider indeed (always in an isotropic context, with diagonal pressure)
a model of sources performing a transition
from an equation of state that is
typical of string-driven pre-big-bang, $p=-\r/d$ (see Section 3), to the
dual, radiation-dominated regime with $p=\r/d$. We shall model this
transition by
$$
{\pb\over \rb}=\ga(\xi)={\xi\over d\sqrt{\xi^2+\xi_1^2}}~~~~,~~~~
\int \ga d\xi={1\over d}\sqrt{\xi^2+\xi_1^2} , \eqno(5.31)
$$
where $|\xi_1|$ is a phenomenological parameter (typically of order $M_P
^{-1}$) that characterizes the time scale of the transition regime.
Asymptotically, i.e. for $|\xi|>>|\xi_1|$, one recovers radiation for
$\xi>0$  and the dual state for $\xi<0$.

By inserting this matter behaviour into the equations
(5.14), (5.15) and (5.20), we get a regular solution provided
$\xi_0$ and $\xi_1$ are chosen such that
$$
\a c>\xi_0^2 , \eqno(5.32)
$$
where $\a$ and $c$ are the constants defined in eq. (5.22). This
condition can be satisfied, in particular, by the choice
$$
\xi_0=0~~~~~~~,~~~~~~~4V_0=\xi_1^2 , \eqno(5.24)
$$
which nicely simplifies the final expression for the background fields
(a more general choice does not change the qualitative behaviour of the
solution, which is presented here for illustrative purposes only).

With this choice, the integration of eqs. (5.15) and (5.16) gives
$$
e^{\fb}=e^{\phi_0}(1+{\xi^2\over \xi_1^2})^{{-d\over d-1}}
\eqno(5.25)
$$
$$
a=a_0({\xi\over \xi_1}+\sqrt{1+{\xi^2\over \xi_1^2}})^{{2\over d-1}}
\eqno(5.26)
$$
where $a_0$ and $\phi_0$ are integration constants. From their
combination we have
$$
e^\phi=a_0^de^{\phi_0}(1+{\xi\over \sqrt{\xi^2+\xi_1^2}})^{{2d\over
d-1}} . \eqno(5.27)
$$
Moreover, according to eq. (5.14),
$$
{\rb \over \r_0}={d-1\over 4d}\xi_1^2\r_0 e^{\phi_0}
(1+{\xi^2\over \xi_1^2})^{-{1\over d-1}}
={d\xi\over dt} \eqno(5.28)
$$
so that
$$
e^\phi
{\r \over \r_0}={d-1\over 4d}\xi_1^2\r_0 e^{\phi_0}
(1+{\xi^2\over \xi_1^2})^{-{d+1\over d-1}}
 \eqno(5.29)
$$
and
$$
e^\phi
{p \over \r_0}={\xi\over \xi_1}{d-1\over 4d}\xi_1^2\r_0 e^{\phi_0}
(1+{\xi^2\over \xi_1^2})^{-{3d+1\over 2(d-1)}}
 \eqno(5.30)
$$

We note that, for $\xi\ra \infty$, eq. (5.26) gives
 $a(\xi)\sim \xi^{2/(d
-1)}$, while for $\xi\ra -\infty$, $a(\xi)\sim (-\xi)^{-2/(d
-1)}$.  From the definition (5.8), it thus follows
 that, in the asymptotic
future  $d\xi/dt \sim a^{-1}$, namely $\xi$ tends to coincide with the
conformal time, while in the asymptotic past $\xi$ tends to coincide
with the "dual" time coordinate, as $d\xi/dt \sim a$.

This solution describes a model that is   expanding ($H>0$),  for all
$t$; the Universe, starting from a flat space ($H\ra 0$) and weak
coupling ($e^\phi \ra 0$) regime, evolves through a superinflationary
phase [$a\sim (-t)^{-2/(d+1)}$] dominated by string-like unstable matter
($p=-\r/d$), towards a final decelerated  [$a\sim t^{2/(d+1)}$],
radiation-dominated ($p=\r/d$) phase  with frozen gravitational coupling
($e^\phi=const.$). In $d=3$ spatial dimensions it provides an explicit
realization of the model discussed at the end of Section 3.

The curvature is everywhere bounded, growing from $-\infty$ to $0$, and
decreasing from $0$ to $\infty$. The behaviour of $H$ and $\dot H$ is
given by
$$
H={a^\pr \rb \over a \r_0}={\xi_1\r_0e^{\phi_0} \over 2d}(1+
{\xi^2\over \xi_1^2})^{-{d+1\over 2(d-1)}} \eqno(5.31)
$$
$$
\dot H=H^\pr { \rb \over  \r_0}=-
{(d+1)\xi \xi_1\r_0^2e^{2\phi_0} \over 8d^2}(1+
{\xi^2\over \xi_1^2})^{-{3d+1\over 2(d-1)}} . \eqno(5.32)
$$
The absence of singularity can be traced back to the fact that this
solution is self-dual, in the sense that
$$
[{a(-t)\over a_0}]^{-1}={a(t)\over a_0}. \eqno(5.33)
$$
In general relativity, such a solution is not allowed, as the field
equations are time-symmetric but not invariant under the inversion of
the scale factor. The behaviour of $a,e^\phi,e^\phi \r,e^\phi p, H$ and
$\dot H$ is plotted in {\bf Fig. 3} for $d=3$,
$\phi_0=0$, $a_0=\r_0=\xi_1=1$.

An analogous solution obviously exists in which $H$ is always negative
(the universe is always isotropically contracting); it is easily
obtained in correspondence with the dual equation of state $\ga(\xi)=
-\xi/d\sqrt{\xi^2+\xi_1^2}$. More interesting, in our context, is
however the Bianchi I type anisotropic background in which, during the
pre-big-bang phase, $d$ dimensions expand with scale factor $a$, while
$n$ dimensions shrink with scale factor $b=a^{-1}$, with an equation of
state $p=-q=-\r/(d+n)$ (the example discussed in Section 3).

In this case, by setting $\ga=p/\r$, we have
$$
{1\over 2}Tr(\Ga\eta)^2=(d+n)(\int \ga d\xi)^2 , \eqno(5.34)
$$
and the equations (5.15) and (5.16) for $a$ and $\fb$ reduce to
$$
\eqalign{
{\fb}^{\pr}&=-{2\over D}(\xi+\xi_0) \cr
{a^\pr \over a}&={2\over D}\int \ga d\xi =-{b^\pr \over b}, \cr}
\eqno(5.35)
$$
where
$$
D=4V_0+(\xi+\xi_0)^2-(d+n)(\int \ga d\xi)^2 . \eqno(5.36)
$$
In analogy with the previous example, we represent the evolution in time
of the sources between the dual asymptotic regimes by
$$
\eqalign{
{p\over \r}&=\ga(\xi)={1\over d+n}{\xi\over \sqrt{\xi^2+\xi_1^2}}=-
{q\over \r} \cr
\int \ga d\xi &={\sqrt{\xi^2+\xi_1^2} \over d+n} \cr}
\eqno(5.37)
$$
and we choose conveniently the arbitrary parameters in such a way that
$\xi_0=0$ and $\xi_1^2=4V_0$.

The integration of eqs. (5.35) then provides
$$
e^{\fb}=e^{\phi_0}(1+{\xi^2\over \xi_1^2})^{-{d+n\over d+n-1}}
\eqno(5.38)
$$
$$
a=b^{-1}=
a_0({\xi\over \xi_1}+\sqrt{1+{\xi^2\over \xi_1^2}})^{{2\over d+n-1}}
\eqno(5.39)
$$
with $a_0$ and $\phi_0$ arbitrary integration constants. It follows that
$$
e^\phi=a_0^{d-n}e^{\phi_0}({\xi\over \xi_1}+\sqrt{1+{\xi^2\over \xi_1^2
}})^{{2(d-n)\over d+n+1}}
(1+{\xi^2\over \xi_1^2})^{-{d+n\over
d+n-1}} \eqno(5.40)
$$
and that
$$
{\rb \over \r_0}={d\xi\over dt}=
{d+n-1\over 4(d+n)}\xi_1^2\r_0 e^{\phi_0}
(1+{\xi^2\over \xi_1^2})^{-{1\over d+n-1}}.
 \eqno(5.41)
$$

By comparing this last equation with eq. (5.39) we can easily check that
 this solution,
in the $\xi\ra -\infty$ limit,  exactly describes the
regime of pre-big-bang and dynamical dimensional reduction discussed in
Section 3, with $a(t)\sim (-t)^{-2/(d+n+1)}=1/b$. This background,
according to the solution (5.39), (5.40), evolves
 in such a way as to reach
a phase of maximal, finite curvature, after which it approaches the
dual, decelerated regime in which the internal dimensions are not
frozen, but keep contracting like $b(t)=1/a \sim t^{-2/(d+n+1)}$ for
$t\ra +\infty$.

It is important to stress that the dilaton coupling, in this case,
 does not settle down to a finite constant value after the big-bang, but,
according to eq.(5.40),   tends to decrease during the phase of
decreasing curvature. Such a decrease of $\phi$ is driven by the
decelerated shrinking of the internal dimensions which are not frozen,
unlike in the previous case.

We have thus provided two examples of self-dual solutions which can
model the transition from the growing to the decreasing curvature
regime. These examples are not intended to apply too far away from the
maximum curvature scale (i.e. in  too low a curvature regime where, for
example, there are constraints which
rule out  too fast a variation of the
gravitational coupling [49] and of the radius of the internal dimensions
[50], such as the ones predicted by our last example).
In the  low curvature regime $O(d,d)$ symmetry and scale
factor duality may indeed
be expected  to be broken, for instance by a dilaton potential
$V(\phi)$.
They suggest,
however, the interesting possibility of a transition, which may occur
even for monotonically evolving scale factors, and which approximates a
de Sitter phase in the neighbourhood of the big-bang region ($\dot H=0$).

We note, finally, that in the model of sources that we have used the
ratio $p/\r$ is not a constant, but the pressure is still diagonal. The
effective, time-dependent equations of state (5.31) and (5.37) may
 thus be
re-expressed in terms of an effective bulk viscosity, which depends on
$\r$ and which becomes negligibly small in the "in" and "out" asymptotic
regimes where the perfect fluid behaviour is recovered. It is thus
significative, in this context, that the evolution of a phase of
exponential de Sitter expansion down to the decelerated regime, in a
model of string-driven inflation [51], can also be interpreted in terms
of an effective $\r$-dependent, bulk viscosity [52].

We have already seen, on the other hand, that in the pre-big-bang phase
the entropy of the sources is constant, as the matter evolution is
adiabatic. Near the maximum ($\dot H\simeq 0$), however, one can define
a horizon entropy [53] which shrinks like the pre-big-bang horizon area.
Viscosity, and other dissipative processes, may thus naturally be expected
to occur, in that regime, possibly yielding a large entropy
increase.

\vskip 1.5cm
{\bf 6. Conclusions}

In this paper we have described how string theory suggests  and
  supports  a picture of the Universe in which the present decelerated
expansion is preceeded by a dual phase in which the evolution is
accelerated and the curvature is growing (what we have called
{\bf "pre-big-bang"}). We stressed that such a scenario is not to be
regarded as an alternative to the usual post-Planckian cosmology, whether
  inflationary or not, but that it represents {\bf a complement} of the
standard scenario, which cannot be extended beyond the
Planck scale.

In the pre-big-bang phase the growth of curvature may be accompanied
 (but not necessarily sustained) by the shrinking of some extra "internal"
spatial dimensions  down to a final compactification scale. This scale,
as well as the   maximal (big-bang)
curvature scale, are both expected to be given by the fundamental
length parameter of string theory, i.e. roughly by the Planck scale
itself. In the
subsequent decelerated phase the internal radius may be frozen, or may
keep shrinking, as shown by the examples presented in Section 5.

As we   pointed out there, the possibility of avoiding
the big-bang singularity appears to be deeply rooted in a typical
stringy symmetry, the scale factor duality,
 which acts on homogeneous cosmological backgrounds.
Indeed, the product of scale-factor-duality
 and time reversal admits, as non-trivial fixed points,
backgrounds which connect smoothly, at $t=0$, two duality-related
regimes, e.g. a superinflationary era and a standard
 decelerating expansion.

Moreover, the generalization of scale-factor-duality to a full
continuous $O(d,d)$ symmetry provides a framework
 in which entire classes of non-singular
cosmological models can be obtained
via $O(d,d)$ "boosts" of trivial, or even singular [45]
initial conformal backgrounds.

Obviously, the picture presented here is in many respects
qualitative and preliminary. Many problems remain to be
solved, such as that of formulating a simple, yet consistent
equation of state for string sources in curved backgrounds.
At a more fundamental level, we might expect that, unless
higher-order quantum string effects are consistently taken
into account, it will be impossible to avoid a singularity
once all physical constraints are imposed.

Nevertheless, we wish to stress that the global picture
emerging from our simplified approach is already  well
defined and allows, in principle, for experimental verifications.
We have shown, indeed, that the  string cosmology
equations and the assumption of a pre-big-bang era provide definite
predictions for the spectrum of   relic gravitons  that should be
filling up space not less than  the electromagnetic CMB does.
 Thus such a scenario can  be confirmed,
disproved, or at least significantly constrained, by some (hopefully
 near-future) direct or indirect observations and measurements
 of a gravitational-wave
   background of cosmological origin. It is perhaps encouraging
 to recall at this point that the-hot big-bang hypothesis [54] was
definitively confirmed  through the
observation of the relic electromagnetic background radiation [55].
\vskip 2 cm
{\bf Acknowledgements}

One of us (MG) wishes to thank J. Ellis and the CERN Theory Division for
hospitality and financial support during part of this work.
\vfill\eject

\centerline{\bf References}
\vskip 1 cm
\item{1.}E.Alvarez, Phys.Rev.D31(1985)418.

\item{2.}Y.Leblanc, Phys.Rev.D38(1988)3087.

\item{3.}R.Brandenberger and C.Vafa, Nucl.Phys.B316(1989)391.

\item{4.}E.Alvarez and M.A.R.Osorio, Int.J.Theor.Phys.28(1989)949.

\item{5.} M.Gasperini, N.Sanchez and G.Veneziano, Int.J.Theor.Phys.A6
(1991)3853;

Nucl.Phys.B364(1991)365.

\item{6.}M.Gasperini and G.Veneziano, Phys.Lett.B277(1992)256.

\item{7.}E.Alvarez, J.Cespedes and E.Verdaguer, Phys. Lett.
 B289 (1992) 51;

A.Ashtekar, "Emergence of discrete structures at the Plank scale", to

appear in Proc. of the 6th Int. Workshop on Theor. Phys. "String quantum

gravity and physics at the Planck energy scale" (Erice, June 1992).

\item{8.}A.Vilenkin, "Did the universe have a beginning?" Caltech preprint

CALT-68-1772 (1992).

\item{9.}F.Lucchin and S.Matarrese, Phys.Lett.B164(1985)282.

\item{10.}G.Veneziano, Phys.Lett.B265(1991)287.

\item{11.}A.A.Tseytlin, Mod.Phys.Lett.A6(1991)1721

\item{12.}A.A.Tseytlin and C.Vafa, Nucl.Phys.B372(1992)443;

A.A.Tseytlin, Class. Quantum Grav.9(1992)979.

\item{13.}K.A.Meissner and G.Veneziano, Phys.Lett.B267(1991)33;

 Mod.Phys.Lett.A6(1991)3397.

\item{14.}A.Sen, Phys.Lett.B271(1991)295;

S.F.Hassan and A.Sen, Nucl.Phys.B375(1992)103.

\item{15.}D.Shadev, Phys.Lett.B137(1984)155;

R.B.Abbott, S.M.Barr and S.D.Ellis, Phys.Rev.D30(1984)720;

E.W.Kolb, D.Lindley and D.Seckel, Phys.Rev.D30(1984)1205.

\item{16.}M.Gasperini, Phys.Lett.B258(1991)70;
 Gen.Rel.Grav.24(1992)219.

\item{17.}N.Sanchez and G.Veneziano, Nucl.Phys.B333(1990)253.

\item{18.}H.J.De Vega and N.Sanchez, Phys.Lett.B197(1987)320.

\item{19.}Nguyen S.Han and G.Veneziano, Mod.Phys.Lett.6(1991)1993.

\item{20.}G.Veneziano, Helvetica Physica Acta 64 (1991) 877.

\item{21.}S.Weinberg, Gravitation and cosmology   (Wiley, New York, 1971)
Chapt.15.

\item{22.}A.Giveon, Mod.Phys.Lett.A6(1991)2834.

\item{23.}N.D.Birrell and P.C.W.Davies, Quantum fields in curved space
  (Cambridge Univ. Press, Cambridge, 1982) Chapt.3.

\item{24.}L.Parker, Nature 261(1976)20

\item{25.}E.Verdaguer, "Gravitational particle creation in the early
Universe",

Barcelona preprint, UAB-FT-276 (Lectures given at the ERE-91 Meeting,

 Bilbao, 1991).

\item{26.}A.H.Guth, Phys.Rev.D23(1981)347.

\item{27.}A.D.Linde, "Inflation and quantum cosmology",
preprint CERN-TH.5561/89

(based on lectures given at the Summer School on Particle Physics and

Cosmology, Trieste, 1989).

\item{28.}L.P.Grishchuk, Sov.Phys.Usp.31(1988)940.

\item{29.}L.P.Grishchuk and M.Solokhin, Phys.Rev.D43(1991)2566.

\item{30.}K.S.Thorne, in 300 Years of Gravitation", eds.  S.W.Hawking
and W.Israel

 (Cambridge Univ.Press, Cambridge, 1988) p.330.

\item{31.}E.M.Lifshitz and I.M.Khalatnikov, Adv. Phys.12(1963)208.

\item{32.}L.P.Grishchuk, Sov.Phys.JEPT 40(1975)409.

\item{33.}L.H.Ford and L.Parker, Phys.Rev.D16(1977)1601.

\item{34.}M.Gasperini and M.Giovannini, "Dilaton contributions
 to the cosmic
gravitational wave background", Turin Univ.  preprint, DFTT 58/92
(to appear in Phys. Rev. D).

\item{35.}J.Garriga and E.Verdaguer, Phys.Rev.D39(1989)1072;

M.Demianski, in Proc.   9th Italian Conference on General Relativity,

 Capri, 1990, eds.  R.Cianci et al. (World Scientific, Singapore,
1991) p.19;

M.Amendola, M.Litterio and F.Occhionero, Phys.Lett.B237(1990)348.

\item{36.}M.Gasperini and M.Giovannini, Class. Quant. Grav.9(1992)L137.

\item{37.}B.Allen, Phys.Rev.D37(1988)2078.

\item{38.}V.Sahni, Phys.Rev.D42(1990)453.

\item{39.}M.Gasperini and M.Giovannini, Phys.Lett.B282(1992)36.

\item{40.}D.Shadev, Phys.Rev.D39(1989)3155;

R.G.Moorhouse and J.Nixon, Nucl.Phys.B261(1985)172.

\item{41.}V.A.Rubakov, M.V.Sazhin and A.V.Veryaskin, Phys.Lett.B115
(1982)189;

R.Fabbri and M.D.Pollock, Phys.Lett.B125(1983)445;

L.F.Abbot and M.B.Wise. Nucl.Phys.B244(1984)54.

\item{42.}G.Smoot et al., COBE collaboration
 preprint 92-04 (to appear in Ap.J.Lett.)

\item{43.}L.Krauss and M.White, "Grand unification, gravitational
waves and the CMB anisotropy",Yale preprint YCTP-P15-92;

 T. Souradeep and V.Sahni, "Density perturbations, gravity
waves and the

 cosmic microwave background", preprint IUCAA
(Pune, India), July 1992;

F.Lucchin, S.Matarrese and S.Mollerach, "The gravitational wave

contribution
to CMB anisotropies
and the amplitude of mass fluctuations

from COBE results",
Fermilab-Pub-92/185-A (July 1992).

\item{44.}L.P.Grishchuk and Y.V.Sidorov, Class. Quantum Grav.6(1989)L161;

Phys.Rev.D42(1990)3413.

\item{45.}M.Gasperini, J.Maharana and G.Veneziano, Phys.Lett.B272(1991)
277;

M.Gasperini, J.Maharana and G.Veneziano, "Boosting away singularities

from conformal string backgrounds", CERN-TH.6634/92;

A. Giveon and A. Pasquinucci, " On cosmological string backgrounds with

 toroidal isometries, Princeton preprint, IASSNS-HEP-92/55.

\item{46.}A.A.Tseytlin, Phys.Rev.Lett. 66 (1991) 545.

\item{47.}A.A.Tseytlin, "String cosmology and dilaton",  to appear in

Proc. of the 6th Int. Workshop on Theor. Phys. "String quantum gravity

and physics at the Planck energy scale" (Erice, June 1992).

\item{48.}V.A.Belinski, E.M.Lifshitz and I.M.Khalatnikov, Adv.Phys.31
(1982)639.

\item{49.}R.W.Hellings, "Time variation of the gravitational constant",
in Proc.   10th Int. School on Cosmology and Gravitation, Erice
1987, ed.   V.N.Melnikov and V. De Sabbata (Kluwer Acad. Pub.,
Dordrecht) p.215;

F.S.Accetta, L.M.Krauss and P.Romanelli, Phys.Lett.B248(1990)146.

\item{50.}E.W.Kolb, M.J.Perry and T.P.Walker, Phys.Rev.D33(1986)869;

J.D.Barrow, Phys.Rev.D35(1987)1805.

\item{51.}N.Turok, Phys.Rev.Lett.60(1987)549.

\item{52.}J.D.Barrow, Nucl.Phys.130(1988)743.

\item{53.}P.C.W.Davies, Class.Quantum Grav.4(1987)L225.

\item{54.}G.Gamow, Phys.Rev.70(1946)572;

R.A.Alpher and R.C.Herman, Rev.Mod.Phys.22(1950)153;

R.H.Dicke, P.J.E.Peebles, P.G.Roll and D.T.Wilkinson, Ap.J.142(1965)414.

\item{55.}A.A.Penzias and R.W.Wilson, Ap.J.142(1965)419.

\vfill\eject

Insert tables here
\vfill\eject
\centerline{\bf Figure captions}

\vskip 2 cm
\noi
{\bf Fig. 1}

\noi
Two possible alternatives to the curvature singularity of the standard
cosmological model.
\vskip 2 cm
\noi
{\bf Fig. 2}

\noi
Present constraints on the maximum allowed curvature scale $H_1$ (in
units of Planck mass), versus the index $\nu$ which parametrizes the
background kinematics. The allowed region extends down to
$|\nu + 1/2|=0$.

\vskip 2 cm
\noi
{\bf Fig. 3}

\noi
Time evolution with respect to $\xi$
of the self-dual solution given by eqs. (5.26) and (5.27) around
the phase of maximal curvature ($\xi =0$).
 a) The scale factor and the dilaton
coupling; b) the effective density and pressure; c) the Hubble parameter
and its time derivative.
\vfill\eject

\end